\documentclass[runningheads]{llncs}
\usepackage{graphicx}
\usepackage{amsmath}
\usepackage{amsfonts}
\usepackage{amssymb}
\usepackage{amsfonts}
\DeclareMathOperator{\sign}{sgn}
\newcommand{\subf}[2]{%
  {\small\begin{tabular}[t]{@{}c@{}}
  #1\\#2
  \end{tabular}}%
}
\begin{document}
\title{On Steering Swarms}
%\title{On Steering Swarms\thanks{This research was partly supported by Technion Autonomous Systems Program (TASP)}}
\author{Ariel Barel\orcidID{0000-0003-3275-4264} \and
Rotem Manor\orcidID{0000-0002-2504-1509} \and
Alfred M. Bruckstein}
\authorrunning{A. Barel et al.}
\institute{Technion - Israel Institute of Technology, Technion City, Haifa 32000, Israel
\index{Barel, Ariel}
\index{Manor, Rotem}
\index{Bruckstein, Alfred M.}
%Princeton University, Princeton NJ 08544, USA \and
%Springer Heidelberg, Tiergartenstr. 17, 69121 Heidelberg, Germany
\email{arielbarel@gmail.com}\\}
%\url{http://www.cs.technion.ac.il}\\}
% \and
\maketitle              % typeset the header of the contribution
\begin{abstract}
The main contribution of this paper is a novel method allowing an external observer/controller to steer and guide swarms of identical and indistinguishable agents, in spite of the agents' lack of information on absolute location and orientation. Importantly, this is done via simple global broadcast signals, based on the observed average swarm location, with no need to send control signals to any specific agent in the swarm.

%\keywords{Steering \and Guiding \and Control \and Multi-Agent \and Decentralized.}
\end{abstract}

%% ---- Bibliography ----
%%
%% BibTeX users should specify bibliography style 'splncs04'.
%% References will then be sorted and formatted in the correct style.
%%
 %%\bibliographystyle{splncs04}
%% \bibliography{mybibliography}
%Author, F.: Article title. Journal \textbf{2}(5), 99--110 (2016)
%Author, F., Author, S.: Title of a proceedings paper. In: Editor,
%F., Editor, S. (eds.) CONFERENCE 2016, LNCS, vol. 9999, pp. 1--13.
%Springer, Heidelberg (2016). \doi{10.10007/1234567890}
%Author, F., Author, S., Author, T.: Book title. 2nd edn. Publisher,
%Author, A.-B.: Contribution title. In: 9th International Proceedings
%on Proceedings, pp. 1--2. Publisher, Location (2010)
%LNCS Homepage, \url{http://www.springer.com/lncs}. Last accessed 4
%\end{thebibliography}

\section{Introduction}\
This paper deals with steering multi-agent systems, based on decentralized gathering laws, using an external broadcast control signal. Agents move according to local information provided by their sensors. The agents are assumed to be identical and indistinguishable, memoryless (oblivious), with no explicit communication between them. The agents do not share a common frame of reference i.e. agents are not equipped with either GPS systems or compasses. By assumption, agents sense the distance and/or bearing to their neighbours, within a finite or infinite range of visibility. An external observer/controller continuously monitors the swarm's location and broadcasts the same control signal, based on the centroid of the agents' constellation. We present a simple yet practical method to steer the swarm and guide it  to a given destination.

Note that unlike the simple agents that are anonymous, unaware of their position, lack memory, and do not use explicit communication to maintain the swarm cohesion, the external controller does need the ability to continuously monitor the trajectory of the swarm location. Due to these capabilities, the controller is able to influence the movement of the swarm, with a very simple global control signal broadcast simultaneously to all agents.
%\section{The inspiration}\label{SteeringInspiration}

The inspiration to this control method came from the following observation: some of the gathering algorithms, while they ensure the convergence of agents to a bounded area, do not imply that the centroid of the agents' location remains stationary in the plane \cite{ando1999,jadbabaie2003,olfati2006flocking,ji2007,olfati2007consensus,gordon2008,bellaiche2015,manor2016Chase,BarelSimpleAgents}. In fact, some gathering algorithms exhibits random walk like behaviour of the centroid of the agents' constellation after gathering as discussed in \cite{Barel2016Come}. The method to steer the swarm to a target point, presented herein, exploits the movements of the system's center of gravity due to the agents' compliance with the distributed convergence algorithm.
\section{How to Control a Single Agent}\label{Single}
We first describe the basic idea in conjunction with a single agent performing a random walk in the plane, and then extend the discussion to multi-agent systems carrying out various cohesion ensuring gathering algorithms. Assume a drunkard agent is moving in the plane in the following random way: at discrete times $k=1, 2, 3, ...$ he selects a new destination for time $k+1$. The destination location $\tilde{p}(k+1)$ is randomly and homogeneously distributed in a unit disc centered at its current position $p(k)$, so that $\tilde{p}(k+1)=p(k)+\tilde{\Delta}(k)$, where $\tilde{\Delta}(k)$ is a random vector uniformly distributed in a unit disc. After selecting $\tilde{p}(k+1)$ the agent starts going there from $p(k)$ in a straight path. By monitoring his motion, one can steer him in any direction with the following control rule: if the projection of his current movement on the required direction is positive - allow the drunkard to finish his step. Otherwise, stop him after a fraction of the unit interval $\mu<1$, by broadcasting (shouting) a startling ``stop!" signal.

This process will cause the drunkard to perform a biased walk, making, in expectation, bigger steps in the desired direction. To bring the drunkard toward a region near a precise target point in the plane, one may define the desired direction to always point from the current location of the drunkard to the goal. Assume first, for simplicity, that the desired direction is fixed.
Let $p(k)$ be the current position of the agent and let $d \in \mathbb{R}^2$ be a unit vector in the direction in which we require the agent to move. Denote by $\tilde{\Delta}(k)$ the \textit{planned} travel vector of the agent for the current time period $[k, k+1)$, from $p(k)$, its position at time $k$, to a homogeneously distributed random point in a unit disc centered at $p(k)$, and by $\Delta(k)$ its \textit{actual} travel vector. The relation between $\tilde{\Delta}(k)$ and $\Delta(k)$ is as follows : at time $k$ the agent starts traveling from its existing position $p(k)$ to its planned position $\tilde{p}(k+1)=p(k)+\tilde{\Delta}(k)$ in a piecewise constant velocity equal to $\tilde{\Delta}(k)/1$. If $\tilde{\Delta}(k)^Td \leq 0$, the external controller stops the agent at a fraction $\mu$ of the time-step, i.e. $\Delta(k)=\mu\tilde{\Delta}(k)$, otherwise the controller does not interrupt its motion during the current time period, hence $\Delta(k)=\tilde{\Delta}(k)$. Therefore we have

\begin{equation} \label{AgentDynamics}
\begin{split}
p(k+1) = p(k)+c(k)\tilde{\Delta}(k)\\
c(k) =\left\{ 
\begin{array}{ll}
\mu &\quad\quad \tilde{\Delta}(k)^Td < 0 \\
1 &\quad\quad o.w.
\end{array}
\right.
\end{split}
\end{equation}
where $\tilde{\Delta}(k)$ is a vector from $p(k)$ to the homogeneously distributed random point in a unit disc centered at $p(k)$.
By symmetry of the random distribution function, for any direction $x$, we have that the expectation of a planned step is $\mathbf{E}\{\tilde{\Delta}x(k)\}=0$.
The required direction of movement $d$ is, without loss of generality, towards the positive $x$ axis, i.e. to the right. Clearly, by the symmetry of the distribution function, we have that the probabilities that the drunkard moves right and left are same and equal $0.5$. Hence, the expected actual travel of the agent, given external controller's (possible) interruptions, is (omitting the time index $(k)$ for simplicity):
\begin{equation}\label{ExpectedJumpSingleStart}
\mathbf{E}\{\Delta x \}=0.5 \mathbf{E}\{\Delta x \mid \tilde{\Delta} x \geq 0 \}+ 0.5 \mathbf{E}\{\Delta x \mid \tilde{\Delta} x < 0\}=0.5(1-\mu)\mathbf{E}(\tilde{\Delta}x \mid \tilde{\Delta}x \geq 0)
\end{equation}
In order to guide an agent to a target point, the controller can set the required direction at each time-step, from the current position of the agent to the target point. Let us find the expected position of the agent at time $(k+1)$ given $p(k)$, i.e. $\mathbf{E}\{\|p(k+1)\|^2 \mid p(k)\}$. By the law of cosines in a triangle \cite{Barel2018Steering} we obtain that
\begin{equation} \label{ABC}
\mathbf{E}\{\|p(k+1)\|^2 \mid p(k)\}=p(k)^2-A(\frac{1-\mu}{2})\|p(k)\|+B(1+\mu^2)
\end{equation}
where
$
A=\mathbf{E}\left\{\frac{\tilde{\Delta}(k)^T p(k)}{\|p(k)\|}\sign\left\{\frac{\tilde{\Delta}(k)^T p(k)}{\|p(k)\|}\right\}\right\}
$
is positive and depends only on the direction vector $d(k)=\frac{p(k)}{\|p(k)\|}$, and for a rotationally symmetric $\tilde{\Delta}(k)$ it is independent of $d(k)$ (and on $p(k)$ of course), and
$
B=\mathbf{E}\{\|\tilde{\Delta}(k)\|^2\}
$
is positive and obviously independent on $p(k)$. From this result it follows that
\begin{equation} \label{ABCfinal}
\mathbf{E}\{\|p(k+1)\|^2\}=\mathbf{E}\{\|p(k)\|^2\}-\left(A\left(\frac{1-\mu}{2}\right) \mathbf{E}\{\|p(k)\|\}-B(1+\mu^2)\right)
\end{equation} 

\begin{figure}%[!htbp]
%\captionsetup{width=0.8\textwidth}
  \centering
       \includegraphics[width=\textwidth]{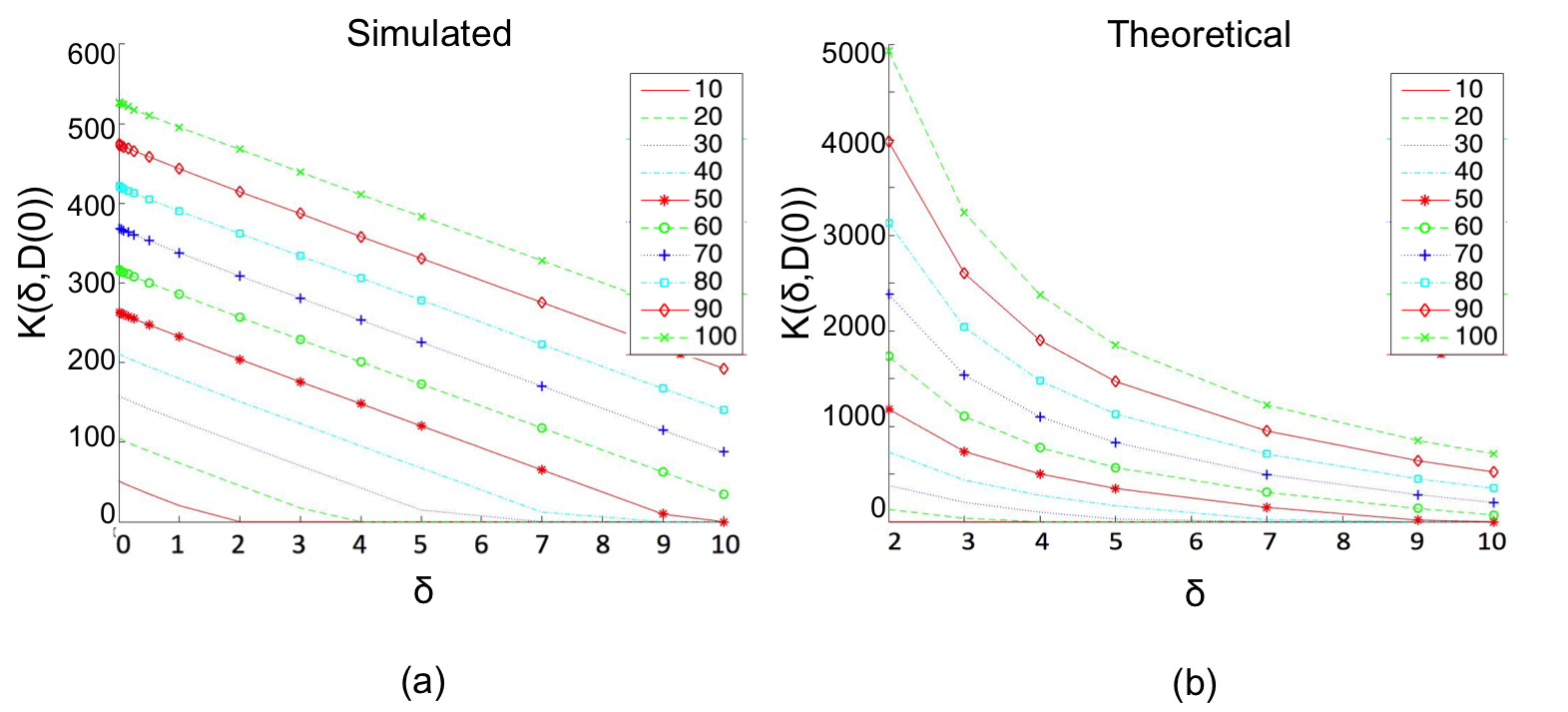}
    \caption{Plot of $k$ vs. $\delta$ for some $D(0)$ values from $10$ to $100$ units. (a) Simulation results, and (b) The theoretical bound. Here $\mu=0.1$ and number of simulation runs is $10,000$.}
      \label{SteeringTheoryVsSimulated}
\end{figure}

We have that if the right expression in big parentheses in (\ref{ABCfinal}) is bigger than $\delta$, $\mathbf{E}\{\|p(k)\|^2\}$ decreases by $\delta$, and while this inequality persists, it will decrease until $\mathbf{E}\{\|p(k)\|\} \leq \left(\frac{B(1+\mu^2)+\delta}{A(\frac{1-\mu}{2})}\right)$. Returning to (\ref{ABCfinal}) we have that after $k(\delta)$ steps, given by
\begin{equation}\label{k_vs_D0}
k(\delta)=\frac{  D^2(0)- \left(\frac{B(1+\mu^2)+\delta}{A(\frac{1-\mu}{2})}\right)^2}   {\delta}
\end{equation}
the process will necessarily stop and the agent will be ``near" the target.
Simulated results of $k$ vs. $\delta$ for some different initial values of $D(0)$ and the graph of Equation (\ref{k_vs_D0}) plotted in Figure \ref{SteeringTheoryVsSimulated} shows that the theoretical $k(\delta)$ is indeed a rather loose upper bound on the number of steps needed to reach the target's neigbourhood.
%\clearpage

\section{Controlling Multi-Agent Systems - the Idea}
Let us adopt this steering method to a multi-agent system. Suppose there is a multi-agent system which converges to a bounded area. The lack of a global orientation of the agents prevents the viewer from simply broadcasting the desired direction of movement as suggested by Azuma et. al. \cite{azuma2013broadcast} and others, since the agents are unable to obey global-direction-based commands. Research methods that draw inspiration from animal behaviour in herds in nature e.g. \cite{couzin2005effective} are based on the fact that part of the group moves in a certain direction and indirectly influences the group's behaviour, but in this article we assume that even leaders do not know how to orient themselves and find the desired direction of movement. Additionally, recall that our agents are anonymous and indistinguishable, hence an external observer wishing to lead the system in a required direction can not steer individual agents separately by transmitting control commands to each one of them. We show here that an external observer can lead a multi-agent system in a required direction (while the agents also converge to a bounded region), by only sensing the motion of the system's centroid. This information represents for the external controller the location of the group, and it is feasible to measure or estimate in real life multi-agent scenarios, especially for large numbers of agents, such as swarms of drones. Let $p_{cm}(t) = \frac{1}{n}\sum_{j=1}^{n} p_i(t)$ be the system's centroid. The velocity of the centroid is the average velocities of the agents 
$\dot{p}_{cm}(t)=\frac{1}{n}\sum_{j=1}^{n}\dot{p}_i(t)$
and we have that while all agent velocities are constant the centroid velocity is constant as well. We assume that during each time interval $k=1, 2, 3, ...$ each agent's velocity is constant, therefore we have that $\hat{\dot{p}}_{cm}(t)$, the direction of the centroid movement is piecewise constant (i.e. does not change during time intervals hence moves in straight lines).
Similar to our discussion in section \ref{Single}, here, the external controller tracks the motion of the \textit{centroid} of the system. If the projection of its movement is on the required direction ($\tilde{\Delta}_{cm}(k)^Td\geq 0$) - it allows \textit{all the agents} to finish their planned travels. Otherwise, it stops them all after a fraction $\mu$ of the time-step, i.e. when they complete a fraction $\mu$ their planned travel. We discuss in detail different types of such systems, and bound the expected ``velocity" of the swarm's centroid due to this control mechanism.
\subsection{Steering a System of Agents with Infinite Visibility and Full Sensing}\label{S2}
We begin with a simple linear multi-agent gathering process in discrete time for the infinite visibility and full sensing case. Each agent $i$ moves according to the decentralized dynamic law: $p_i(k+1)=p_i(k)-\sigma\sum_{j=1}^{n}(p_i(k)-p_j(k))$, where $0 < \sigma < \frac{2}{n}$ is a constant gain factor, i.e. at each time-step, each agent jumps proportionally to the sum of relative position vectors to all the other agents (recall system $\mathcal{S}_2$, in \cite{Barel2016Come}). As proved by Gazi, Passino et. al. \cite{gazi2004stability}, since the dynamics of such system is governed by an antisymmetric pairwise interaction function, the average position of the agents is invariant.
To steer this system in some desired direction, we would like to bias the motion of the system centroid by measuring its trend, hence we assume some additive ``noise" that breaks symmetry and causes the center of the system to move. We hence assume that each agent, in addition to obeying the distributed control law above, also moves to a randomly selected point at each time step:
\begin{equation}
p_i(k+1)=p_i(k)-\sigma\sum_{j=1}^{n}(p_i(k)-p_j(k))+\tilde{\Delta}_i(k)
\label{eq:Dynamics2_1_With_noise}
\end{equation}
where $\tilde{\Delta}_i(k)$ is a randomly selected point in a unit disc. Here too, at time $k$ the agents start traveling from their existing positions $p_i(k)$ towards their next planned positions $\tilde{p}_i(k+1)$ in piecewise constant velocities equal to their distance from it $[-\sigma\sum_{j=1}^{n}(p_i(k)-p_j(k))+\tilde{\Delta}_i(k)]/1$, so that if an external controller does not intervene, all the agents arrive at their destinations simultaneously at time $k+1$. Hence we may denote the planned motion of the centroid to be $\tilde{\Delta}_{cm}(k)= \bar{\tilde{p}}(k+1)-\bar{p}(k)=\frac{1}{n}\sum_{i=1}^{n}\tilde{\Delta}_i(k)$, and the control mechanism for system (\ref{eq:Dynamics2_1_With_noise}) is:
\begin{equation} \label{MultiAgentDynamics}
\begin{aligned}
\begin{split}
&p_i(k+1) = p_i(k)+c(k)[-\sigma\sum_{j=1}^{n}(p_i(k)-p_j(k))+\tilde{\Delta}_i(k)]\\
&c(k) =\left\{ 
\begin{array}{ll}
\mu &\quad\quad \tilde{\Delta}_{cm}(k)^Td < 0 \\
1 &\quad\quad o.w.
\end{array}
\right.
\end{split}
\end{aligned}
\end{equation}

Here $c(k)$ represents the optional ``stop" signal received simultaneously at fraction $\mu$ of the time-step by all agents, $\tilde{\Delta}_{cm}(k)=\frac{1}{n}\sum_{i=1}^{n}\tilde{\Delta}_i(k)$ is the planned travel of the centroid of the agents, and $d$ is the required direction of movement of the system. Since the projection on $x$ of the second moment of a disc of radius $r$ is $\frac{1}{4}\pi r^4$, we have in this system \cite{Barel2018Steering} that $\mathbf{E}\{\Delta x_{cm}\} \geq 0.5 (1-\mu)\frac{1}{8n}$ i.e. the bound on the expected step of the centroid is inversly proportional to the number of agents. To guide a system to a goal point, the observer controller should set the desired direction at every time interval so $d(k)$ is a unit vector from the centroid of the system to the goal point. Figure \ref{Simulations} presents a typical simulation result of this system with full visibility and complete sensing, with some evenly distributed noise jump to a unit disc of each agent, as presented in equation (\ref{MultiAgentDynamics}).

\subsection{Steering a System of Agents with Limited Visibility and Bearing Only Sensing}\label{Manor}
\begin{figure}[ht]
\centering
\begin{tabular}{|c|c|}
\hline
\subf{\includegraphics[width=59mm]{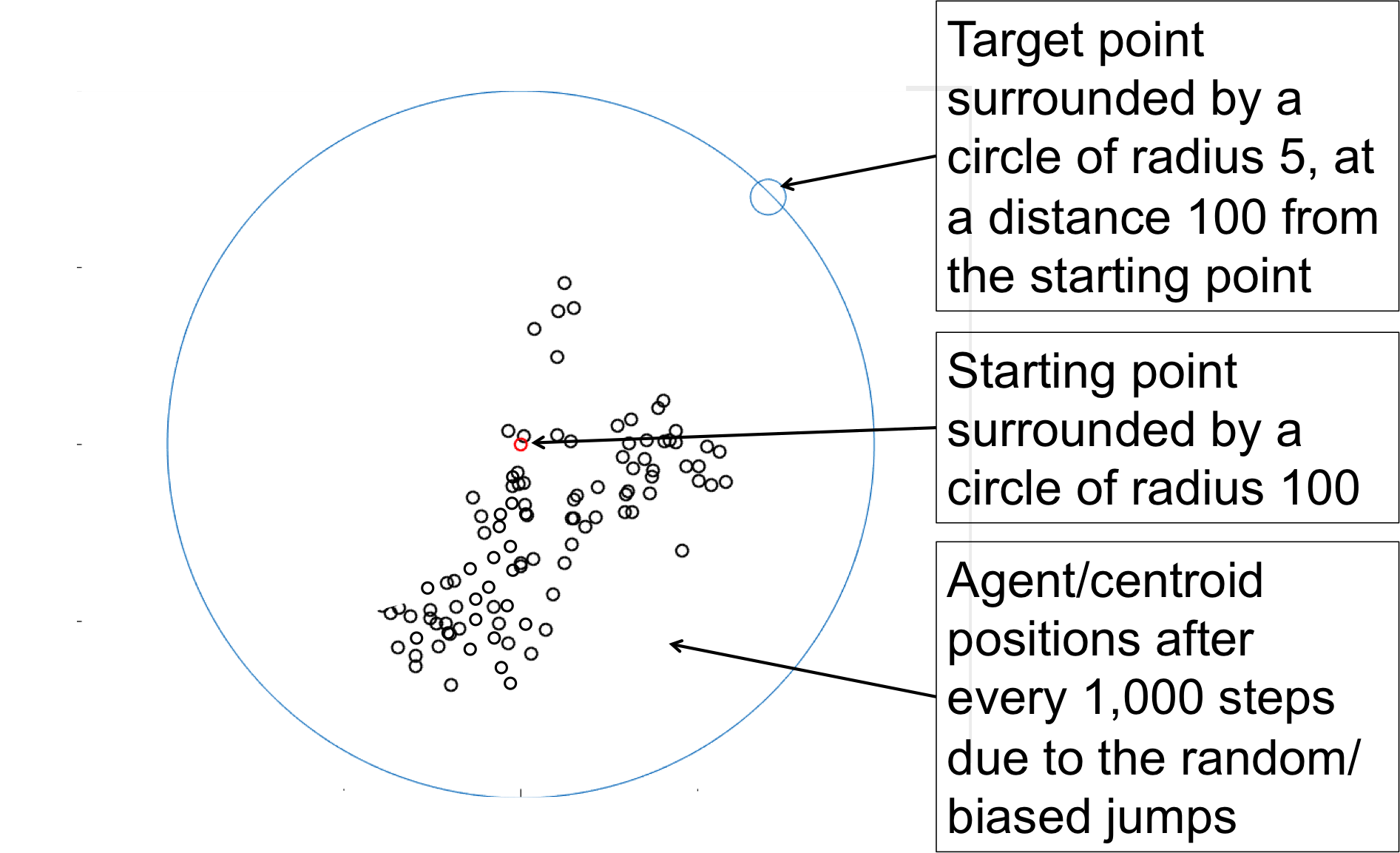}}
     {(a)}
&
\subf{\includegraphics[width=59mm]{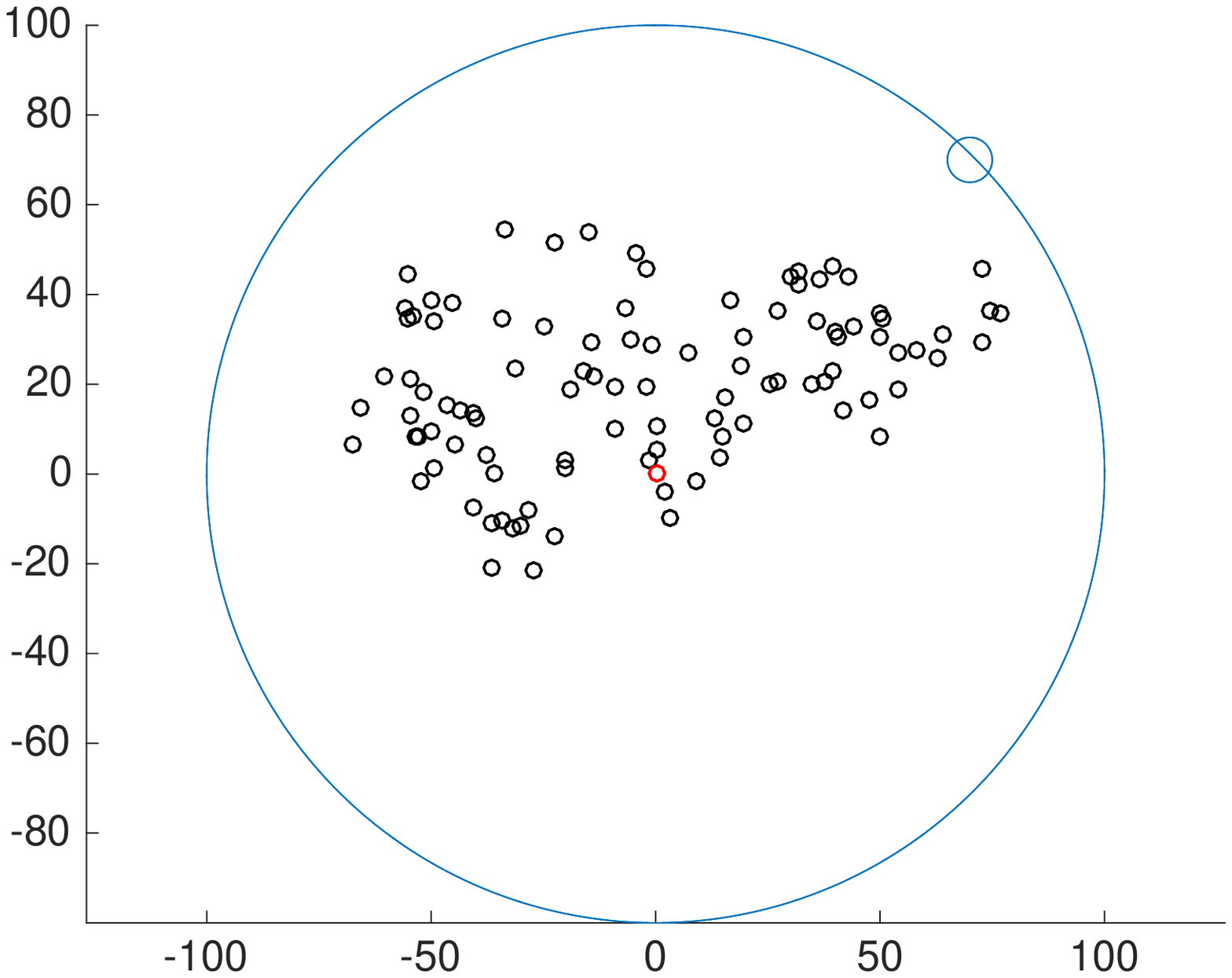}}
     {(b)}
\\
\hline
\subf{\includegraphics[width=59mm]{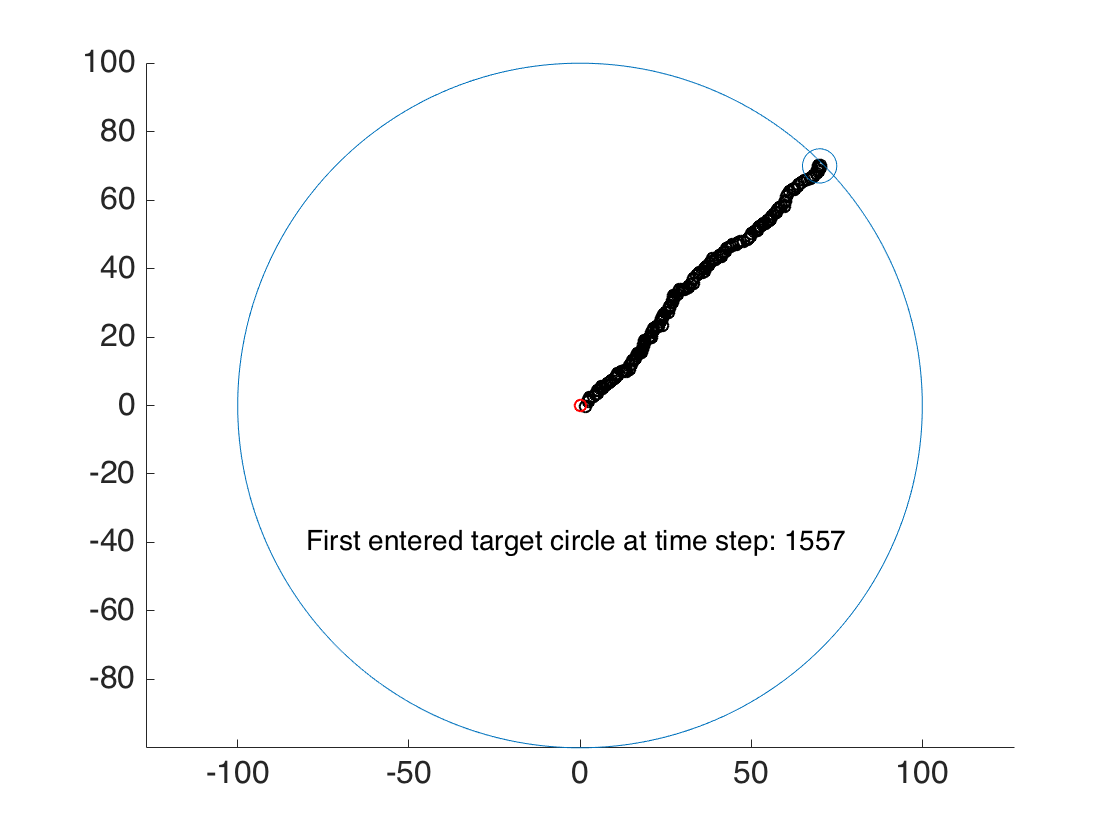}}
     {(c)}
&
\subf{\includegraphics[width=59mm]{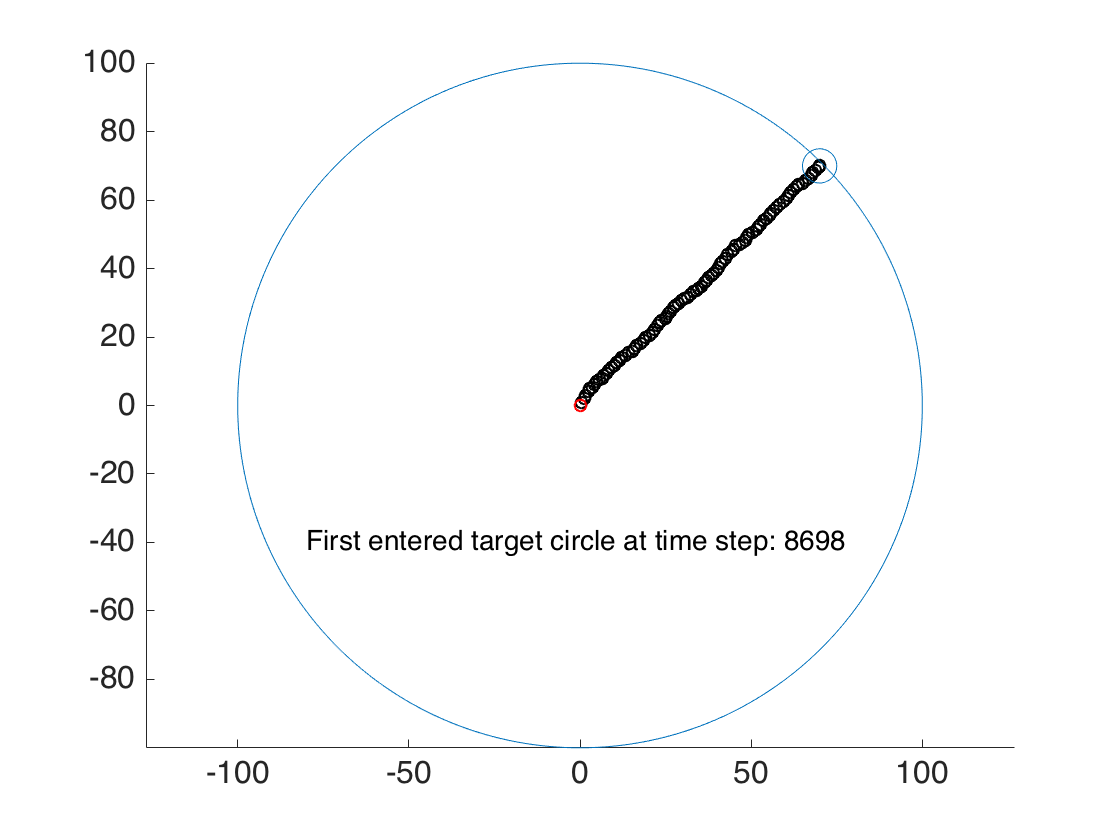}}
     {(d)}
\\
\hline
\end{tabular}
\caption{Random walk vs. Steering a multi-agent system to a goal point: (a) General legend of the simulation settings. (b) Typical $100,000$ random unit steps of a drunkard agent with no bias (agents' position was plotted every $1,000$ steps for enhanced readability). (c) Typical simulation run of the system in section \ref{S2} with $n=10$ and $\mu=0.01$. The system centroid first entered the goal area in less than $1,600$ time steps. (d) Typical simulation run of the system in section \ref{Manor} with $n=10$ and $\mu=0.01$. The system centroid first entered the goal area in less than $9,000$ time steps.}
\label{Simulations}
\end{figure}

Here we assume that the agents are able to sense the direction to their neighbours (i.e. bearing only sensing), and their motions being determined by the set of unit vectors pointing from their current location to their neighbours. The neighbours are defined for each agent $i$ at time-step $k$ as the set of agents located within a given visibility range $V$ form its position $p_i(k)$. Manor et. al. \cite{manor2016Bear} modified Gordon's et. al. motion laws \cite{gordon2004} \cite{gordon2005}, and proved that the new law gathers the agents of the system to a disc with a radius equal to the agents' maximal step size $\sigma$ within a finite expected number of time steps, and that the distribution of the agents' average position converges in probability to the distribution of a random-walk.
As in section \ref{S2}, we assume here piecewise continuous dynamics (where agents continuously move towards their new locations), so that the formal steering algorithm for this system is:
\begin{equation} \label{BO_RandomDynamicsWithDirectionControl}
\begin{aligned}
\begin{split}
&p_i(k+1) =\left\{ 
\begin{array}{ll}
 p_i(k) &\quad \psi_i(k) \ge \pi \mbox{ or } \chi_i(k) = 0 \\
p_i(k)+c(k)\tilde{\Delta}_i(k) &\quad o.w.\\
\end{array}
\right.\\
&\chi_i(k) =
\left\{
\begin{array}{ll}
1 & \quad\quad \mbox{w.p. } \delta \\
0 &\quad\quad  \mbox{w.p. } 1-\delta
\end{array}
\right.\\
&c(k) =\left\{ 
\begin{array}{ll}
\mu & \quad\quad \tilde{\Delta}_{cm}(k)^Td < 0 \\
1 & \quad\quad o.w.\\
\end{array}
\right.\\
&\tilde{\Delta}_i(k)=\mbox{vector from $p_i(k)$ to a random point in $ar_i(k)$}\\
\end{split}
\end{aligned}
\end{equation}
where $\tilde{\Delta}_{cm}(k)=\sum_{i=1}^{n} \tilde{\Delta}_{i}(k)$ is the planned jump of the centroid of the system, and $d$ is a unit vector in the required moving direction of the system. It was proved in \cite{manor2016Bear} that the original model, given no external control, satisfies
$
\mathbf{E}\{\Delta_{cm}(k)\} =0
$, and that
\begin{equation}\label{StepSizeRightManor}
\mathbf{E}\{\Delta x_{cm}\} \geq 0.25 (1-\mu)\frac{1}{n^2}Var^*
\end{equation}
$$Var^*=\delta^2(\frac{\sigma}{2})^2\frac{1-\cos^4(\frac{\pi-\psi_*}{2})}{\frac{\pi-\psi_*}{2}-\frac{1}{2}\sin(\pi-\psi_*)}$$
Figure \ref{Simulations} presents simulations result of this system (\ref{BO_RandomDynamicsWithDirectionControl}). The system gathers and moves to a goal, and the trace of the travel of the system's centroid is plotted.
%\newpage
%\subsection{Simulation results}
%************************************************************************
%\begin{figure}[h]
%\begin{center}$
%\begin{array}{ll}
%\includegraphics[width=60mm]{RandomJumpsExplenation}
%\includegraphics[width=60mm]{100Keach1000_5}\\
%\includegraphics[width=60mm]{CirclesS2_10AgentsMu001PlotEach10Steps_6}
%\includegraphics[width=60mm]{CirclesManor_10AgentsMu001PlotEach10Steps_5}
%%\includegraphics[width=70mm]{50Keach1000_9}&
%%\includegraphics[width=70mm]{50Keach1000_10}
%\end{array}$
%\end{center}
%\caption{Random walk vs. Steering a multi-agent system to a goal point. Figure $(1)$: general legend of the simulation settings. Figure $(2)$: typical random walk of a drunkard agent with no bias (agents' position was plotted every $1,000$ steps for enhanced readability).Figure $(3)$: typical simulation run of the system in section \ref{S2} with $\mu=0.01$. The system centroid first entered the goal area in less than $1,600$ time steps. Figure $(4)$: typical simulation run of the system in section \ref{Manor} with $\mu=0.01$. The system centroid first entered the goal area in less than $9,000$ time steps.}
%\label{Simulations}
%\end{figure}
%************************************************************************

\section{Conclusions}
A method has been introduced here that allows an external observer to control a multi-agent system and guide it to a desired destination even when the agents are very primitive. According to our paradigm all the agents are identical (anonymous), therefore the external observer can not send a separate command to each agent, but can broadcast the same command to all the agents. The viewer controls the swarm by means of an identical command sent simultaneously to all agents.
The method was tested for different cases: the control of a single moving agent performing random-walk, steering of a system with infinite visibility and relative distance and bearing measurement, and control of a system with partial information (limited visibility and bearing only measurement). 

\bigskip
\subsubsection*{Acknowledgments. } 
This research was partly supported by Technion Autonomous Systems Program (TASP).

%\newpage
%\addcontentsline{toc}{section}{References}
%\bibliography{SteeringBib}\
\bibliographystyle{splncs04}
\bibliography{SteeringBib}
\end{document}